\documentclass{elsart1p}
\usepackage{graphicx}
\usepackage{amssymb}

\newcommand{\MUnits}{$GeV/c^2$}
\newcommand{\Bot}{$b$}
\newcommand{\ttbar}{$t\bar{t}$}

\newcommand{\ifb}{$fb^{-1}$}

\begin{document}
\begin{frontmatter}
%
%
%
\title{Top Quark Mass Measurements from CDF}
%
%
\author{Ford Garberson on behalf of the CDF Collaboration}
\address{University of California Santa Barbara Physics, Santa Barbara CA 93106-9530, USA}
\begin{abstract}
This document was written as part of the PANIC 2008 conference proceedings in Eilat Israel.
We review top mass measurements performed at CDF Run II through October of 2008. The basic algorithms used in 
many mass measurements are discussed as are some common systematic
difficulties. Mass results in combination with the D0 Collaboration are presented,
and the implications on the Standard Model Higgs mass are discussed.

\end{abstract}
\end{frontmatter}
\section{\label{sec:template} Mass Measurement Techniques}

There are two categories that most top mass measurements can be grouped into.
Many analyses follow the "template method", in which templates are built for
the signal and background samples which are then used in a maximum likelihood
fit on the data. 

The other major category of analyses follow the "matrix element method". Under this approach one performs mass measurements for each event, and then combines the likelihoods to determine the final result. For a given event, one starts with the measured parton momenta, and then determines the likelihood for a given top mass according to the matrix element calculated from theory. In the end, the matrix element approach uses more information about each event than the template method. Its cost is that it requires extensive integration over
many probability functions, such as the detector resolution functions and the momentum distribution functions of the colliding partons. The latest CDF result in the lepton plus jets
channel, \cite{bib:cdf_lpj_ME}, results in a top mass of $172.2 \pm 1.3(stat+JES)
\pm 1.0(syst)\ GeV/c^2$ using 2.7 \ifb, a resolution of better than one percent on the mass.  

\section{Systematic Uncertainties}
\label{sec:Ident}

Most mass measurements are no longer limited by statistical uncertainties. As a result it becomes very important to determine the systematic uncertainties accurately. Representative examples of systematic uncertainties are shown in Table \ref{dil_lpj_comb_systs} for three template method measurements \cite{bib:cdf_dil_lpj_comb_temp} in the lepton plus jets channel, the dilepton channel, and in a combined measurement. Uncertainties in the simulation of \ttbar\ signal events include
uncertainties on the momenta of the colliding partons, the modeling of hard QCD and QED radiation, and jet fragmentation uncertainties. 

\begin{table}[th]
  \begin{center}
  \caption{Uncertainties for the lepton plus jet, dilepton, and combined channel measurements described in \cite{bib:cdf_dil_lpj_comb_temp}. }\label{dil_lpj_comb_systs}
  \begin{tabular}{|c|c|c|c|}
  \hline
  Uncertainty & LJ & DIL & Both \\
  \hline
  Stats (2.7 \ifb) & 1.7 & 3.0 & 1.6 \\
  Jet Energy Scale & 0.7 & 3.3 & 0.6 \\
  Signal Modeling & 0.8 & 1.5 & 0.8 \\
  Background Modeling & 0.2 & 0.3 & 0.2 \\
  Other & 0.3 & 0.6 & 0.4 \\     
   \hline
  Combined Systs & 1.1 & 3.7 & 1.1 \\
  \hline
  \end{tabular}
  \end{center}
\end{table}

The JES uncertainty is the largest uncertainty on the world average top mass, but two key improvements are expected to greatly reduce it in the future as statistics rise. One of these improvements is to assume all jets in the event have the same jet energy scale and then "calibrate" the scale so that the measured W mass comes out correct for each event. When one applies this calibration,
the top mass is fitted simultaneously with the jet energy scale, and this converts the JES uncertainty from a systematic one to a
statistical one, leaving a residual systematic that is about four times smaller. However, this approach can only be applied when a W boson decays hadronically, which is why the jet energy scale uncertainty is so large in the dilepton channel.

Another approach is to circumvent the jet energy scale uncertainty entirely by
not directly using the jet energy information. CDF has performed two mass
measurements using such variables as well as a combination \cite{bib:cdf_TBM}. The variables used are
the mean transverse momentum of leptons from the W boson decays, and
the mean decay length of the \Bot-jets. With 1.9 \ifb\ at CDF, the
combined measurement has a statistical uncertainty of 6.2 \MUnits\ and a
systematic uncertainty of 3.0 \MUnits. These measurements are currently statistically limited, however at the LHC these measurements should
become directly competitive, and should have minimal correlation to other
measurements in combination.

\section{\label{sec:combination} Top Mass Combination and Higgs Constraints}

A combination of twelve CDF and D0 top mass measurements has recently been
performed \cite{bib:tev_mass_comb}. Since the Standard Model predicts that the Higgs boson will couple with both
the top quark and the W boson, a calculation can be performed to determine
constraints on the Higgs mass given the values and uncertainties on the world
average W mass and top mass. This has been done by the LEP Electroweak working
group \cite{bib:Higgs_constraints}. Accounting for uncertainties on these
masses and on other relevant experimental and theoretical quantities one
determines a constraint on the Higgs mass as shown in Figure \ref{higgs_fig}.
This constrains the Higgs mass to be less than $154\ GeV/c^2$ at a 95\%
confidence level. It should be noted, however, that this limit increases to
$185\ GeV/c^2$ when one accounts for the experimental lower limit determined by
LEP searches. Still, it will be very interesting to see how this limit evolves as the
results of these mass measurements continue to improve.

\begin{figure*}
\centerline{
  \mbox{\includegraphics[height=2.5in]{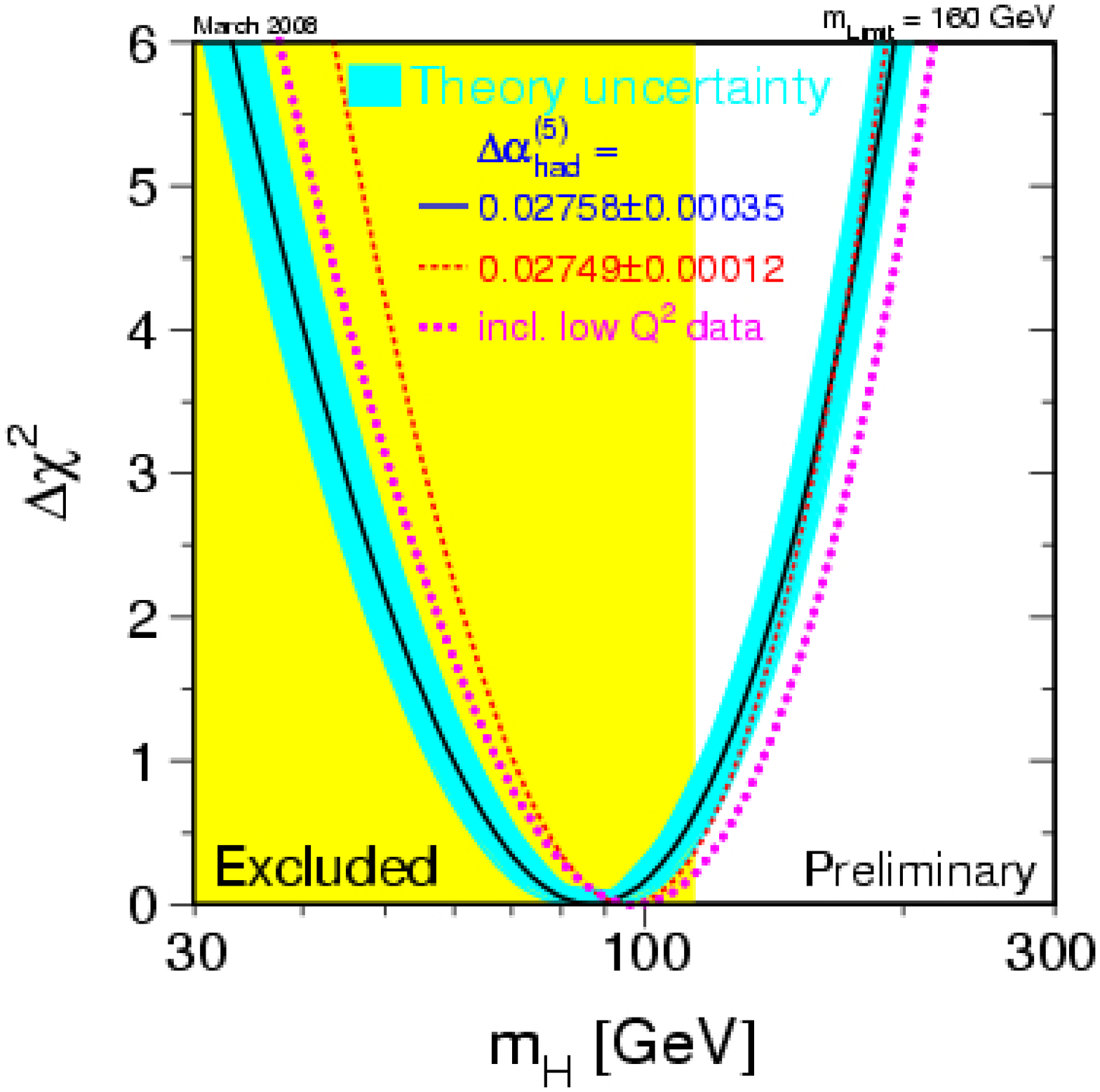}}
  \mbox{\includegraphics[height=2.5in]{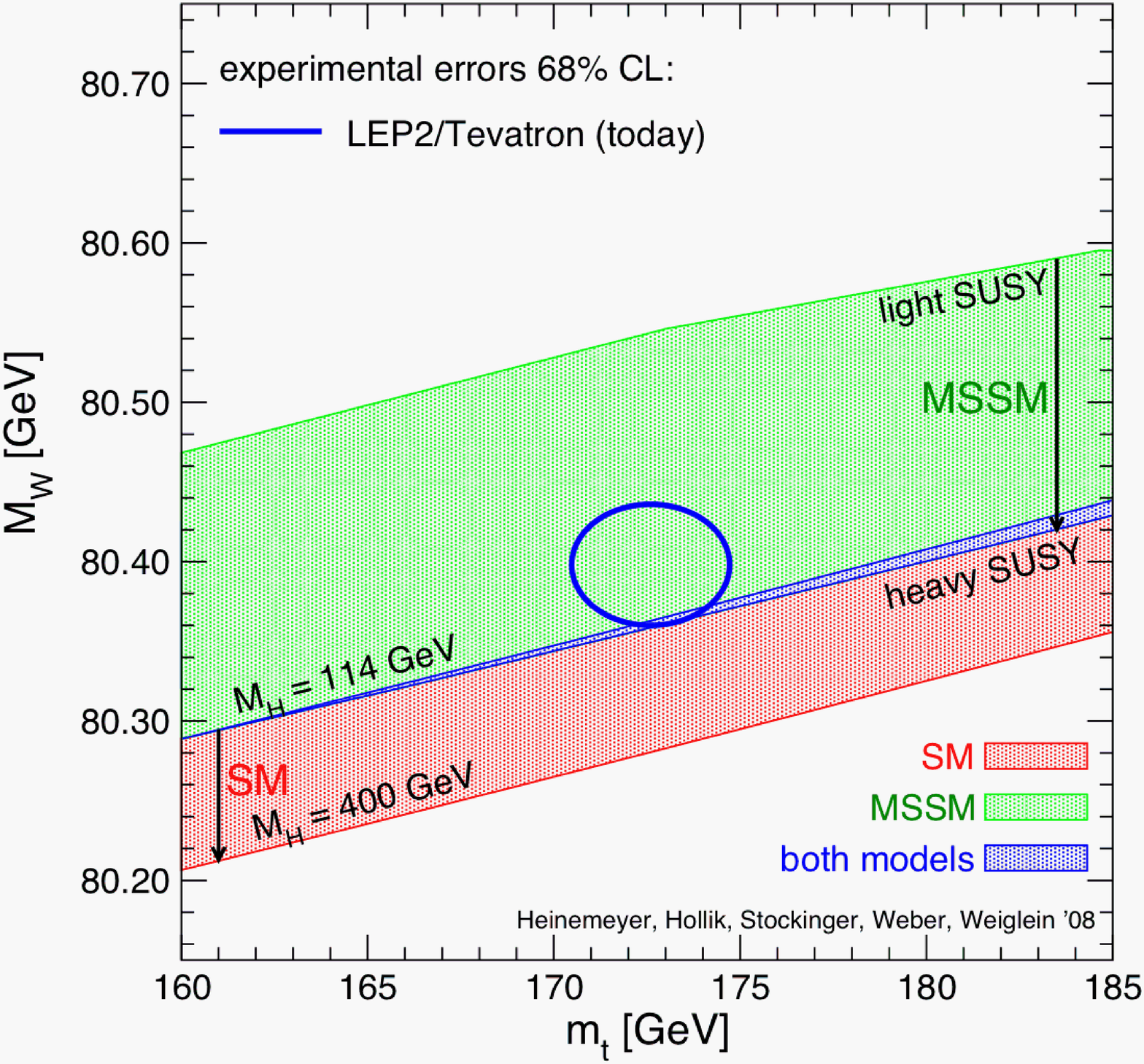}}
}
\caption{Left: One sigma Higgs mass constraints given the world average top and W mass results. Right: The likelihood distribution for the Higgs mass, calculated by the LEP Electroweak Working Group. The black distribution is used to determine the 95\% confidence intervals listed in the text. See \cite{bib:Higgs_constraints}. }
\label{higgs_fig}
\end{figure*}

\section{\label{sec:conclusion} Conclusion}
The CDF Collaboration has performed numerous measurements of the top mass. In this note there was only space to present a few of them, but many others can be found at the public CDF Top Group webpage \cite{bib:top_webpage}. The world average top mass is no longer statistically limited, and the W mass calibration technique has greatly reduced the dominant jet energy scale uncertainty. Measurements using tracking based variables will reduce these systematics still further at the LHC, but the Tevatron has already determined the mass with very high precision. Using these results and those of the W mass, constraints can now be set on the Standard model Higgs mass which will continue to improve. 



%
%
%

%
\end{document}
%
